\documentclass[
    10pt,
    final,
    journal,
    letterpaper,
    twocolumn,
]{./IEEEtran}
\renewcommand{\baselinestretch}{1.00}
\usepackage{multirow}
\usepackage[dvips]{graphicx}
\usepackage{amsmath}
\usepackage[psamsfonts]{amssymb}
\usepackage{amsxtra}
\usepackage{threeparttable}
\usepackage{subfigure}
\usepackage{cite}
\usepackage{graphicx}
\usepackage{psfrag}
\usepackage{stfloats}
\usepackage{amsmath}
\usepackage{amssymb}
\usepackage{array}
\usepackage{epsfig}
\usepackage{balance}
\usepackage{enumerate}
\usepackage{times}
\usepackage{algorithm}
\usepackage{algorithmic}
\usepackage{color}
\usepackage{epstopdf}
\begin{document}

\title{\LARGE Variable-Length Feedback Codes under a Strict Delay Constraint}
\author{Seong Hwan Kim,~\IEEEmembership{Member,~IEEE},
        Dan Keun Sung,~\IEEEmembership{Fellow,~IEEE},
        and Tho Le-Ngoc,~\IEEEmembership{Fellow,~IEEE}
\thanks{Copyright \copyright 2015 IEEE. Personal use of this material is permitted. Permission from IEEE must be obtained for all other uses, in any current or future media, including reprinting/republishing this material for advertising or promotional purposes, creating new collective works, for resale or redistribution to servers or lists, or reuse of any copyrighted component of this work in other works.}
\thanks{This manuscript has been accepted for publication in IEEE Communications Letters.}
\thanks{S. H. Kim and T. Le-Ngoc are with the Department of Electrical \& Computer Engineering, McGill University, Canada (E-mail: seonghwan.kim@mcgill.ca; tho.le-ngoc@mcgill.ca). D. K. Sung is with the Department of EE, KAIST, Korea (dksung@ee.kaist.ac.kr).}
\thanks{Digital Object Identifier 10.1109/LCOMM.2015.2398866}}

\maketitle
\begin{abstract}
We study variable-length feedback (VLF) codes under a \textit{strict delay constraint} to maximize their average transmission rate (ATR) in a discrete memoryless channel (DMC) while considering periodic decoding attempts. We first derive a lower bound on the maximum achievable ATR, and confirm that the VLF code can outperform non-feedback codes with a larger delay constraint. We show that for a given decoding period, as the strict delay constraint, $L$, increases, the gap between the ATR of the VLF code and the DMC capacity scales at most on the order of $O(L^{-1})$ instead of $O(L^{-1/2})$ for non-feedback codes as shown in Polyanskiy \textit{et al.} [``Channel coding rate in the finite blocklengh regime,'' IEEE Trans. Inf. Theory, vol. 56, no. 5, pp. 2307-2359, May 2010.]. We also develop an approximation indicating that, for a given $L$, the achievable ATR increases as the decoding period decreases.
\end{abstract}
\IEEEpeerreviewmaketitle
\section{Introduction}
\PARstart{T}{he} channel capacity, i.e., the maximum coding rate with
an arbitrarily small error probability, needs an assumption of
infinite block-length~\cite{Thomascover}. However, the block error probability (BEP) is non-zero in practice since the block-length must be finite.
For given block-length and BEP constraints, lower and upper bounds on the maximum coding rate were found in ~\cite{IOP6}. Meanwhile, feedback is known to be very useful for performance improvement if variable-length codes are applied.
In the regime of asymptotically long average block-length, the error exponent, regarded as the exponential rate of decay of the BEP with respect to the average block-length, has been an important performance measure for variable-length codes with feedback \cite{Burnashev, Burnashev2, Yamamoto, Shulman, Draper, Tchamkerten, Tchamkerten2}. However, the analysis cannot present the maximum coding rate for fixed BEP and fixed average block-length.

Recently, Polyanskiy et al.\cite{Polyanskiy2} focused on the regime of fixed BEP and fixed average block-length in their study of variable-length feedback (VLF) codes over discrete memoryless channels (DMC), formulating the following problem:
\begin{align}\label{eq.poly_pro}
\max M\quad \text{s.t.}~{\Pr[W\ne \widehat{W}]\le \epsilon},\quad\mathbb{E}[\tau] \le L,
\end{align}
where $M$, $W$, $\widehat{W}$, and $\tau$ denote the number of codewords, transmitted message, estimated message, and the length of a received sequence. The positive real values $\epsilon$ and $L$ denote constraints. A transmitter sends a codeword as time-domain symbols and a receiver attempts to decode it at every symbol reception and sends a stop-feedback to the transmitter when the receiver decides to decode it. Therefore, $\tau$ also represents a delay. They showed that a significant spectral efficiency gain can be obtained by using VLF codes for given average delay and BEP constraints as compared to the non-feedback fixed-length codes. They also considered a strict delay constraint in formulating the following problem:
\begin{align}\label{eq.poly_pro2}
\max M\quad \text{s.t.}~{\Pr[\{W\ne \widehat{W} \}\cup \{\tau > L\}] = \epsilon}.
\end{align}
Analyzing \eqref{eq.poly_pro2}, they alleged that the strict delay constraint nullifies the gain of VLF codes over the non-feedback fixed-length codes.
However, problem \eqref{eq.poly_pro2} does not maximize the average transmission rate (ATR), $\frac{\log_2{M}}{\mathbb{E}[\tau]}$, which is also defined as the spectral efficiency.
Therefore, it is not certain whether VLF codes have a gain or not over the non-feedback codes in terms of the spectral efficiency, under a strict delay constraint.
Besides VLF codes, they also studied VLF with termination (VLFT) codes in which the transmitter decides to stop transmitting a message by observing the output of the decoder fed back from the receiver and sends a termination signal to the receiver under an assumption of error-free feedback and feed-forward channels.
Chen et al.~\cite{Chen13} added practical constrains to Polyanskiy's VLFT codes by limiting the length of codewords and attempting to decode a codeword only at specified times.
In their setting, the same codeword is reused if the receiver fails to decode it after receiving the whole original codeword while the average delay is limited as given in problem \eqref{eq.poly_pro}. However, the VLFT codes are far from practical because existing feedback codes such as Hybrid Automatic Repeat \& reQuest (HARQ) schemes only use acknowledgement (ACK)/negative-ACK (NACK) feedbacks.

In this letter, we revisit VLF codes under a strict delay constraint and a fixed BEP to maximize the achievable ATR (instead of the number of codewords). The regime of strict (rather than \textit{average}) delay constraints is desired to guarantee delay requirements of real-time traffic applications.
We also consider periodic decoding attempts where the receiver attempts to decode a codeword periodically. We derive the achievable ATR as the lower bound on the maximum ATR of VLF codes. We prove that gap between the ATR of VLF codes and the DMC capacity scales at most on the order of $O(L^{-1})$ for a given decoding period instead of $O(L^{-1/2})$ for non-feedback codes with fixed length of $L$ as shown in \cite{IOP6}. This comparison shows a significant gain of VLF codes over the non-feedback codes under a strict delay constraint. We also derive an approximation of the achievable ATR expression indicating that, for a given $L$, the achievable ATR increases as the decoding period decreases. In~\cite{Shkim_GC}, we studied HARQ schemes under a strict delay constraint in AWGN channels for a given BEP with the following two ideal assumptions: 1) the decoding period is sufficiently long; 2) the NACK event is identical to the error event. In this letter, without the above two assumptions, we find the theoretical bound of VLF codes in DMC.

Throughout this letter, we use the following notation: $X$, $x$, and $P_{X}$ denote a random variable, its sample value, and the probability distribution of $X$, respectively. $x^n=(x_1,x_2,...,x_n)$ denotes an $n$-dimensional vector and $x_j$ the $j$th element of $x^n$. $\log(x)=\log_e(x)$ unless otherwise stated.

\section{Problem Statement}\label{SEC.Prob_Statemet}
In this section, we briefly introduce the channel and VLF codes which are modified from those in \cite{Polyanskiy2} for defining periodic decoding attempts, and formulate our optimization problem.

\textit{Channel:} A DMC consists of a pair of input $X$ and output $Y$ on the finite alphabets $\mathcal{A}$ and $\mathcal{B}$, respectively, with a conditional probability, $P_{Y_i|X_i} = P_{Y_1|X_1}, \forall i \ge 1$.

\textit{$(l,d)$ VLF code:} An $(l,d)$ VLF code with $M$ messages, maximum allowable number of decoding attempts, $l$, and decoding period (i.e., interval between 2 consecutive decoding attempts), $d$, is defined as follows:
\begin{enumerate}[\quad 1)]
\item A random variable $U\in\mathcal{U}$ with a probability distribution of $P_U$ represents a codebook shared by both transmitter and receiver.
\item A sequence of encoders $f_n: \mathcal{U}\times \{1,\cdots,M\} \rightarrow  \mathcal{A}$ represent the channel input at time $n$, $X_n = f_n(U,W)$ where $W \in \{1,\cdots,M \}$ is the equi-probable message.
\item A sequence of decoders $g_k:\mathcal{U}\times\mathcal{B}^{dk} \rightarrow \{1,\cdots,M\}$ attempting to provide the estimate of $W$ at time $dk$ where $k$ denote the number of decoding attempts, respectively.
\item A final decision is made by the receiver at a stopping time $d \tau^*$ : $\widehat{W} = g_{\tau^*}(U,Y^{d \tau^*})$.
\end{enumerate}

\textit{Optimization problem:}
For an $(l,d)$ VLF code with $0<\epsilon<1$, we maximize its achievable ATR under a strict delay constraint as follows:
\begin{align}\nonumber
\mathcal{T}^*_{f}(l,d,\epsilon) = \mathop{\max}\limits_{M} \frac{\log M}{d\mathbb{E}[\tau^*]}\quad \text{s.t.}~\begin{array}{*{20}l}
   {\Pr[W\ne \widehat{W}]\le \epsilon,}  \\
   {\Pr[\tau^*\le l]=1.}  \\
\end{array}
\end{align}
$L=dl$ represents the limited length of the received sequence (i.e., the delay). Setting $d=1$ indicates the case of attempting to decode a codeword at every symbol reception.

\section{Achievability Analysis}\label{SEC.per_decodng}
For achievability analysis, we specify the codebook, encoder, and decoder which are modified from those in [10] by adding functions to satisfy strict delay constraints as follows:

\textit{Codebook:} A codebook $U$ is defined on space $\mathcal{U}$ such as
\begin{align}\nonumber
\mathcal{U}&\triangleq \underbrace{\mathcal{A}^{dl}\times \cdots \times \mathcal{A}^{dl}}_{M{\rm{ times}}},\quad P_U\triangleq \underbrace{P^{dl}_{X} \times \cdots \times P^{dl}_{X}}_{M{\rm{ times}}},
\end{align}
where $X$ is distributed according to $P_{X}$ on $\mathcal{A}$.
The realization of $U$ defines $M$ $dl$-dimensional vectors $\mathbf{C}_j \in \mathcal{A}^{dl},j=1,\cdots,M$.

\textit{Encoder:} The encoding sequence $f_n$ maps an equi-probable message $j$ to $\mathbf{C}_j \in \mathcal{A}^{dl}$ and provides the channel input at time $n$,
\begin{align}\nonumber
f_n(j) = (\mathbf{C}_j)_n\quad\text{for}~1\le n \le dl,
\end{align}
where $(\mathbf{C}_j)_n$ is the $n$th coordinate of the vector $\mathbf{C}_j$.

\textit{Decoder:}  A decoder computes the $j$-th information density at the $k$-th decoding attempt for $1\le k \le l$,
\begin{align}
S_{j,k} \triangleq i(\mathbf{C}_{j}(dk),Y^{dk}),\quad j=1,\cdots,M,\nonumber
\end{align}
where $\mathbf{C}_j(n)$ is the first $n$ symbols of $\mathbf{C}_j$ and, the information density between $x^n$ and $y^n$ is defined as
$$i(x^n ;y^n ) = \log \frac{{dP_{Y^n |X^n } (y^n |x^n )}}{{dP_{Y^n } (y^n )}}.$$ The decoder defines $M$ stopping instances computing the information densities
\begin{align}\nonumber
\tau_j \triangleq \left\{ {\begin{array}{*{20}l}
   { \inf\{0\le k\le l: S_{j,k}\ge \gamma \}} & {\text{if}\quad  \mathop{\vee}\limits_{k=1}^{l} \{S_{j,k}\ge \gamma\}} \\
   {l+1} &{{\text{Otherwise,}}} \\
\end{array}} \right.
\end{align}
where $\gamma$ is the threshold to stop computing information density for each codeword, and for statements $\mathsf{A}$ and $\mathsf{B}$, $\mathsf{A} \lor \mathsf{B}$ (resp., $\mathsf{A} \land \mathsf{B}$) is true if $\mathsf{A}$ or $\mathsf{B}$ (resp., $\mathsf{A}$ and $\mathsf{B}$) are true. Since $\mathop{\wedge}\limits_{k=1}^{l} \{S_{j,k}<\gamma\}$ can be checked at time $l$, a mapping $\tau_j=l+1$ does not mean that the observation at the $(l+1)$-th decoding attempt is required. Define a random variable
\begin{align}\label{eq.tau_appo}
\tau' \triangleq \min\{\tau_1,\cdots,\tau_M\}.
\end{align}
The final decision is made by the decoder at the stopping time
\begin{align}\label{eq.tau_star}
\tau^* \triangleq \left\{ {\begin{array}{*{20}l}
   {\tau'} & {\text{if}\quad \tau'\le l} \\
   {l} &{\text{if}\quad \tau' = l+1,} \\
\end{array}} \right.
\end{align}
i.e., $\tau^*$ is always smaller than or equal to $l$.
The output of the decoder is
\begin{align}\label{eq.output_decoder}
g(Y^{d\tau^*})
\triangleq \left\{ {\begin{array}{*{20}l}
   {\max\{j:\tau_j=\tau'\}} & {\text{if}\quad \tau'\le l} \\
   {M} &{\text{if}\quad \tau' = l+1.} \\
\end{array}} \right.
\end{align}

For the VLF code specified above, we find a lower bound on $\mathcal{T}^*_{f}(l,d,\epsilon)$.

\textit{Lemma 1:} For an arbitrary DMC with capacity $C$, interval $d$, and $\exp\{-Cdl/2\}<\alpha<\epsilon$, the maximum $M$ satisfying ${\Pr[W\ne \widehat{W}]\le \epsilon}$ and $\tau^* \le l$ is lower-bounded by
\begin{align}\nonumber
\mathop {\max M}\limits_{\scriptstyle \Pr [W \ne \hat W] \le  \epsilon \hfill \atop
  \scriptstyle \tau^*  \le l \hfill}
    \! \ge \!\! \left\lfloor (\epsilon-\alpha)\exp\left\{\!\!\left(1-\sqrt{\frac{2\log(1/\alpha)}{Cdl}}\right)\!Cdl\!\right\} \! +1 \!\right\rfloor\!. \end{align}

\textit{Proof:} Let us define the following instances:
\begin{subequations}\label{eq.hitting.time}
\begin{align}\label{eq.hitting.time1}
 \tau &= \inf \{ k \ge 1:i(X^{dk} ;Y^{dk} ) \ge \gamma \},  \\
 \bar{\tau}  &= \inf \{ k \ge 1:i(\bar X ^{dk} ;Y^{dk} ) \ge \gamma \},\label{eq.hitting.time2}
\end{align}
\end{subequations}
where $X$ and $\bar{X}$ are independent random variables with the same distribution and $Y$ is the channel output when $X$ is channel input. The average probability of error with $M$ codewords is upper-bounded as
\begin{align}\label{eq.avg_pro_error1}
&\mathbb{P}[g(Y^{d\tau^*})\ne W ]\le \mathbb{P}[g(Y^{d\tau^*})\ne1|W=1]\\\label{eq.avg_pro_error2}
&~= \mathbb{P}\left[ {\left. {\bigcup\limits_{j = 2}^M {\{ \tau _j  \le \tau _1 \le l \} } } \bigcup \{l<\tau_1\} \right|W = 1} \right] \\\label{eq.avg_pro_error3}
&~\le(M-1)\mathbb{P}[\tau_2 \le \tau_1 \le l|W=1] + \mathbb{P}[l<\tau_1|W=1]\\\label{eq.avg_pro_error4}
&~= (M-1) \mathbb{P}[\bar{\tau} \le \tau \le l] +\mathbb{P}[l<\tau]\\
&~\le (M-1) \mathbb{P}[\bar{\tau} \le \tau] +\mathbb{P}[l<\tau],\label{eq.avg_pro_error5}
\end{align}
where \eqref{eq.avg_pro_error1} follows from \eqref{eq.output_decoder}, \eqref{eq.avg_pro_error2} from \eqref{eq.tau_appo}, \eqref{eq.tau_star}, and \eqref{eq.output_decoder}, \eqref{eq.avg_pro_error3} from a union bound and symmetry for $\tau_j$, \eqref{eq.avg_pro_error4} from \eqref{eq.hitting.time}.

Note that, as compared to (45) in \cite{Polyanskiy2}, \eqref{eq.avg_pro_error5} contains the additional term $\mathbb{P}[l<\tau]$ for the event that no codeword is detected within $l$ decoding attempts. $\mathbb{P}[\bar{\tau} \le \tau]$ is upper-bounded by (see (111)-(118) of \cite{Polyanskiy2}\footnote{Polyanskiy et al. proved \eqref{eq.error_detection} only for $d=1$. However, the proof is easily extended to the cases of $d>1$.})
\begin{align}\label{eq.error_detection}
\mathbb{P}[\bar{\tau} \le \tau]\le \exp\{-\gamma\}.
\end{align}
$\mathbb{P}[l<\tau]$ is upper-bounded by
\begin{align}\label{eq.no_detection1}
\mathbb{P}[l<\tau] &= \mathbb{P}\left[\bigcap\limits_{k = 1}^l\{i(X^{dk};Y^{dk})<\gamma\}\right]\\
&\le \mathbb{P}\left[i(X^{dl};Y^{dl})<\gamma\right] \nonumber\\\label{eq.no_detection3}
&=\mathbb{P}\left[ i(X^{dl};Y^{dl})< (1-\delta) Cdl\right]\\\label{eq.no_detection4}
& \le \exp\{-\delta^2Cdl/2\},
\end{align}
where \eqref{eq.no_detection1} follows from \eqref{eq.hitting.time1}, \eqref{eq.no_detection3} is obtained by setting $\gamma = (1-\delta)Cdl$ for $C = \mathbb{E}[i(X;Y)]$ and $0 <\delta < 1$, and \eqref{eq.no_detection4} is obtained by using a Chernoff bound.

Instead of $\mathbb{P}[g(Y^{d\tau^*})\ne W ]\le \epsilon$, we set a stricter constraint using \eqref{eq.avg_pro_error5} and substitute \eqref{eq.error_detection} and \eqref{eq.no_detection4} into \eqref{eq.avg_pro_error5},
\begin{align}\label{eq.error_probability}
(M-1)\exp\{-(1-\delta)Cdl\}+\exp\left\{-{\delta^2 Cdl /2} \right\} \le \epsilon.
\end{align}
Notice that any $M$ satisfying \eqref{eq.error_probability} also satisfies $\mathbb{P}[g(Y^{d\tau^*})\ne W ] \le \epsilon$.
To satisfy \eqref{eq.error_probability}, let $\exp\{-\delta^2Cdl/2\} = \alpha$ for $0<\alpha<\epsilon$. For $0<\delta<1$, the range of $\alpha$ becomes $\exp\{-Cdl/2\}<\alpha<\epsilon$.
Therefore, \eqref{eq.error_probability} can be rewritten as
\begin{align}\nonumber
M\le (\epsilon-\alpha)\exp\left\{\left(1-\sqrt{\frac{2\log(1/\alpha)}{Cdl}}\right)Cdl\right\}+1
\end{align}
for $\exp\{-Cdl/2\}<\alpha<\epsilon$. We know that $\hat{M}(\alpha)=\lfloor (\epsilon-\alpha)\exp\{(1-\sqrt{\frac{2\log(1/\alpha)}{Cdl}})Cdl   \} +1   \rfloor$ is the maximum $M$ satisfying \eqref{eq.error_probability}. In other words, we can achieve $\hat{M}(\alpha)$ with ${\Pr[W\ne \widehat{W}]\le \epsilon}$ and $\Pr[\tau^*\le l]$.\hfill $\blacksquare$

\textit{Lemma 2:} For an arbitrary DMC with capacity $C$, interval $d$, and $\exp\{-Cdl/2\}<\alpha<\epsilon$, $\mathbb{E}[\tau ^* ]$ is upper-bounded as
\begin{align}\nonumber
\mathbb{E}[\tau ^* ] \le {\min \left\{\left(1-\sqrt{\frac{2\log(1/\alpha)}{Cdl}}\right)l+\frac{a_0}{C},l\right\}},
\end{align}
where $a_0$ is the maximum value of $i(X,Y)$.

\textit{Proof:}
\begin{align}\label{eq.avg_tau_star1}
\mathbb{E}[\tau ^* ] & \le \min (\mathbb{E}[\tau' ],l)\\\label{eq.avg_tau_star2}
&\le \min \left( {\frac{1}{M}\sum\limits_{j = 1}^{M} {\mathbb{E}[\tau _j |W = j]} ,l} \right) \\\label{eq.avg_tau_star3}
  &= \min \left( {\mathbb{E}[\tau _1 |W = 1],l} \right) \\\label{eq.avg_tau_star4}
  &= \min \left( {\mathbb{E}[\tau ],l} \right),
\end{align}
where \eqref{eq.avg_tau_star1} follows from \eqref{eq.tau_star}, \eqref{eq.avg_tau_star2} follows from \eqref{eq.tau_appo}, \eqref{eq.avg_tau_star3} follows from symmetry, and \eqref{eq.avg_tau_star4} follows from \eqref{eq.hitting.time1}. Since $i(X^{dk},Y^{dk}) - dkI(X,Y) = i(X^{dk},Y^{dk}) - dkC$ is a martingale, from Doob's optional stopping theorem~\cite{Polyanskiy2}, we obtain $\mathbb{E}[i(X^{d\tau},Y^{d\tau}) - d\tau C] = 0$. Then, $\mathbb{E}[\tau]$ is upper-bounded as
\begin{align}
\mathbb{E}[\tau] &= \frac{\mathbb{E}[i(X^{d\tau},Y^{d\tau})]}{dC} \le \frac{\gamma(\alpha) + da_0}{dC},\label{eq.doobs}
\end{align}
where \eqref{eq.doobs} follows from \eqref{eq.hitting.time1} and $\gamma(\alpha) = \left(1-\sqrt{\frac{2\log(1/\alpha)}{Cdl}}\right)Cdl$. Note that \eqref{eq.avg_tau_star4} is similar to (39) of \cite{Polyanskiy2} except for the limitation to $l$ and \eqref{eq.doobs} is similar to (107) of \cite{Polyanskiy2} except for $d$ and $\gamma(\alpha)$. \hfill$\blacksquare$

\textit{Theorem 1:} For an arbitrary DMC with capacity $C$, and $e^{-\frac{Cdl}{2}}<\epsilon<1$,
\begin{align}\label{eq.th3}
\mathcal{T}^*_{f}(l,d,\epsilon)&\ge \!\!\!\!\max\limits_{e^{-\frac{Cdl}{2}}\!<\alpha<\epsilon} \!\frac{ \log(\epsilon-\alpha)+Cdl-\sqrt{{2Cdl\log(1/\alpha)}}}{d \min \left\{\left(1-\sqrt{\frac{2\log(1/\alpha)}{Cdl}}\right)l+\frac{a_0}{C},l\right\}}.
\end{align}
\textit{Proof:} From Lemmas 1 and 2,
\begin{align}\nonumber
&\max\limits_{M} \frac{\log M}{ d\mathbb{E}[\tau^*|M]} \ge \max\limits_{e^{-\frac{Cdl}{2}}<\alpha<\epsilon} \frac{\log\hat{M}(\alpha)}{d\mathbb{E}[\tau^*|\hat{M}(\alpha)]}\\\nonumber
&\ge\!\!\max\limits_{e^{-\frac{Cdl}{2}}<\alpha<\epsilon}\!\!\!\!\!\! \frac{ \log\left\lfloor(\epsilon-\alpha)\exp\left\{\!\!\left(1-\sqrt{\frac{2\log(1/\alpha)}{Cdl}}\right)Cdl \right\} \!+\!1\right\rfloor}{d\min \left\{\left(1-\sqrt{\frac{2\log(1/\alpha)}{Cdl}}\right)l+\frac{a_0}{C},l\right\}}\\\nonumber
&\ge\max\limits_{e^{-\frac{Cdl}{2}}<\alpha<\epsilon} \frac{ \log(\epsilon-\alpha)+Cdl-\sqrt{{2Cdl\log(1/\alpha)}}}{d\min \left\{\left(1-\sqrt{\frac{2\log(1/\alpha)}{Cdl}}\right)l+\frac{a_0}{C},l\right\}}.
\end{align}
\hfill$\blacksquare$

While Theorem 1 offers the expression to compute the lower bound on the maximum ATR in terms of $C, l,d,$ and $\epsilon$, the expression does not provide a simple view of the effect of $l$ on $\mathcal{T}_f(l,d,\epsilon)$. The following theorem establishes the asymptotic behavior of the gap between $C$ and $\mathcal{T}^{*}_f(l,d,\epsilon)$ as $l$ increases.

\textit{Theorem 2:} For an arbitrary DMC with capacity $C$ and any $e^{-\frac{Cdl}{2}}<\epsilon<1$, the gap between the capacity and the maximum ATR of a VLF code scales at most,
\begin{align}\nonumber
\Delta_{\text{VLF}} \triangleq C - \mathcal{T}^*_{f}(l,d,\epsilon) = O(1/l),
\end{align}
where $f(n) = O(g(n))$ iff there are constants $c$ and $n_0$ such that $f(n)\le c g(n)~\forall n > n_0$.

\textit{Proof:}  $\Delta_{\text{VLF}} = C- \max\limits_{M}\frac{\log{M}}{d\mathbb{E}[\tau^*|M]}$ is upper-bounded as
\begin{align}
&\Delta_{\text{VLF}} \le \!\!\!\!\min\limits_{e^{-\frac{Cdl}{2}}\!<\!\alpha<\epsilon}\!C\!-\!\frac{ \log(\epsilon-\alpha)+Cdl-\sqrt{{2Cdl\log(1/\alpha)}}}{d\min \left\{\left(1-\sqrt{\frac{2\log(1/\alpha)}{Cdl}}\right)l+\frac{a_0}{C},l\right\}}\!\label{eq.delta_upper2}\\
&\le\!\!\!\!\mathop {\min }\limits_{e^{ - \frac{{Cdl}}{2}}  < \alpha  < \min \{\epsilon ,e^{- \frac{{da_0^2 }}{{2Cl}}} \} } {\frac{{da_0  - \log ( \epsilon- \alpha )}}{{\left( {1 - \sqrt {\frac{{2\log (1/\alpha )}}{{Cdl}}} } \right)dl + \frac{{da_0 }}{C}}}},\label{eq.delta_upper4}
\end{align}
where \eqref{eq.delta_upper2} follows from Theorem 1  and \eqref{eq.delta_upper4} follows from limiting the range of $\alpha$ by eliminating $\min$ in the denominator and from the relationship, $\beta_0-\beta_1/\beta_2 = (\beta_2-\beta_1/\beta_0)\beta_0/\beta_2$.
For terms related to $l$ on the right-hand-side (RHS) of \eqref{eq.delta_upper4}, we obtain the following three convergences:
\begin{align}\nonumber
&\mathop {\lim }\limits_{l \to \infty } \exp \left\{-{{Cdl}}/{2} \right\} = 0, \quad \mathop {\lim }\limits_{l \to \infty } \exp \left\{  - {{a_0^2d }}/({2Cl})\right\}=1,\\
&\mathop {\lim }\limits_{l \to \infty }\frac{{ {dl - \sqrt {{{2dl\log (1/\alpha )}}/{{C}}} }  + {da_0}/{C}}}{l}=d.\nonumber
\end{align}
Therefore, there exist constants $l_1$ and $l_2$ such that $\exp \left\{-\frac{{Cdl}}{2} \right\} < \frac{\epsilon}{2}$ for any $l>l_1$ and $\frac{\epsilon}{2} < \exp\{-\frac{a_0^2d}{2Cl}\}$ for any $l>l_2$, respectively. In addition, by setting $\alpha = \epsilon/2$, for a given constant $0<c_0<1$, there exists a constant $l_3>\max\{l_1,l_2\}$ such that $
\frac{{ {dl - \sqrt {\frac{{2dl\log (2/\epsilon )}}{{C}}} }  + \frac{{d a_0 }}{C}}}{l}>c_0d$ for any $l>l_3$. Finally, we have for $l>l_3$
\begin{align}
\Delta_{\text{VLF}} \le {\frac{{da_0  - \log ( 0.5\epsilon )}}{{{dl - \sqrt {\frac{{2dl\log (2/\epsilon )}}{{C}}} }  + \frac{{da_0 }}{C}}}} \le{\frac{{d a_0  - \log ( 0.5\epsilon)}}{l d c_0}}.\label{eq.delta_upper_scale2}
\end{align}
Therefore, $\Delta_{\text{VLF}} = O(1/l)$.\hfill $\blacksquare$\\
Remarks:
\begin{enumerate}[\quad 1)]
\item For given $d$, since $L$ is linearly proportional to $l$, $\Delta_{\text{VLF}}$ scales as
$\Delta_{\text{VLF}} = O(1/L)$ from Theorem 2 .
\item From \eqref{eq.delta_upper_scale2}, the achievable ATR expression is approximated for $L \gg d $ as
\begin{align}
\mathcal{T}^*_f(l,d,\epsilon) \approx C - {\frac{{da_0  - \log ( 0.5\epsilon)}}{L}}.\label{eq.T_approx4}
\end{align}
\item  From \eqref{eq.T_approx4}, for a given $L$, the approximation of $\mathcal{T}^*_f(l,d,\epsilon)$ increases as $d$ decreases.
\end{enumerate}

\section{Illustrative Results}\label{SEC.example}
We show some illustrative results obtained from the analysis in Section \ref{SEC.per_decodng} in terms of ATR [b/s/Hz]. To use \textit{b/s/Hz} as a unit of measure in illustrative results, quantities related with the spectral efficiency are divided by $\log_e 2$.

\begin{figure}
\includegraphics[width=9.0cm]{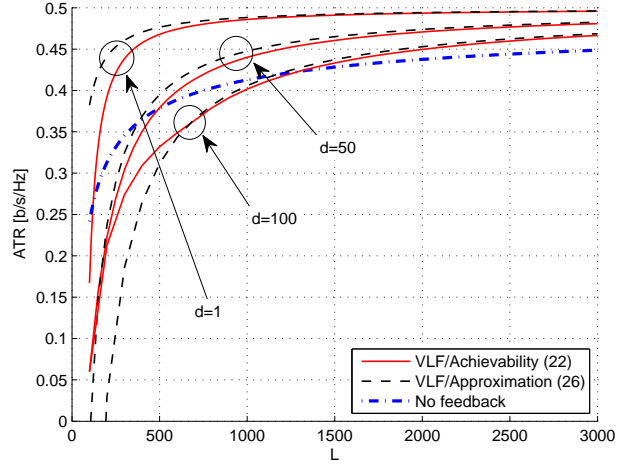}
    \caption{The ATR vs. $L$ for the BSC(0.11); probability of error $\epsilon=10^{-3}$ for $d=1, 50$, and 100. }
    \label{fig.BSC2}
\end{figure}

Fig.~\ref{fig.BSC2} plots the ATR [b/s/Hz] versus $L$ for binary symmetric channel (BSC) with crossover probability $q=0.11$, $\epsilon = 10^{-3}$, and $C=0.5$ b/s/Hz for $d = 1$, 50, and 100. In BSC($q$) channel, $C = \log(2)-H(q)$ and $a_0 = \max\{\log(2q),\log(2(1-q))\}$. The legend `No feedback' corresponds to the fixed blocklength codes without feedback given by Theorem 52 of \cite{IOP6}. It is interesting to observe that the ATR of VLF codes converges to the capacity faster than that of the fixed-length codes even under a strict delay constraint.
When $d=1$, to achieve 90\% of the capacity, $L=360$ is enough for VLF-code while at least 3100 is required for non-feedback code. The ATRs of the VLF code are lower than those of the non-feedback code for $L\le120$ with $d=1$, $L\le450$ with $d=50$, and $L\le1200$ with $d=100$. In VLF codes, decoding attempts at $k<l$ give an opportunity to stop the transmission before $l$. However, these attempts simultaneously increase the probability that other messages than the transmitted one are detected as an estimated message. As $L$ increases, the positive effect of VLF codes dominates the negative effect. Therefore, as $L$ increases, VLF codes offer much higher ATR than non-feedback codes. We can also observe that, as $d$ decreases, the ATR increases for a given $L$ as indicated by \eqref{eq.T_approx4}. In practice, smaller $d$ may
need faster processing or incur higher complexity, which should be considered in the design of VLF codes for a specific application. As shown in Fig.~\ref{fig.BSC2}, the approximation \eqref{eq.T_approx4} provides results very close to those of \eqref{eq.th3}.

\section{\label{SEC:conclu} Conclusions}
We investigated a VLF code under a \textit{strict delay constraint} to maximize the ATR in DMCs while considering periodic decoding attempts. We first derived a lower bound on the maximum achievable ATR and showed that the VLF code can outperform the non-feedback codes for increasing $L$.
We also proved that the gap between the ATR of the VLF code and the DMC capacity scales at most on the order of $O(L^{-1})$ for a given $d$ instead of $O(L^{-1/2})$ for non-feedback codes as previously shown in \cite{IOP6}. The approximated expression of the ATR indicates that the ATR increases with decreasing decoding period $d$.

%
%
\bibliographystyle{./IEEEtran_v111}
\bibliography{./IEEEabrv,./RefAbrv,00_Reference}

\begin{thebibliography}{10}
\providecommand{\url}[1]{#1}
\csname url@rmstyle\endcsname
\providecommand{\newblock}{\relax}
\providecommand{\bibinfo}[2]{#2}
\providecommand\BIBentrySTDinterwordspacing{\spaceskip=0pt\relax}
\providecommand\BIBentryALTinterwordstretchfactor{4}
\providecommand\BIBentryALTinterwordspacing{\spaceskip=\fontdimen2\font plus
\BIBentryALTinterwordstretchfactor\fontdimen3\font minus
  \fontdimen4\font\relax}
\providecommand\BIBforeignlanguage[2]{{%
\expandafter\ifx\csname l@#1\endcsname\relax
\typeout{** WARNING: IEEEtran.bst: No hyphenation pattern has been}%
\typeout{** loaded for the language `#1'. Using the pattern for}%
\typeout{** the default language instead.}%
\else
\language=\csname l@#1\endcsname
\fi
#2}}

\bibitem{Thomascover}
T.~M. Cover and J.~A. Thomas, ``Elements of information theory, 2nd ed.''
  \emph{Wiley-Interscience}, 2006.

\bibitem{IOP6}
Y.~Polyanskiy, H.~V. Poor, and S.~Verdu, ``Channel coding rate in the finite
  blocklength regime,'' \emph{{IEEE} Trans. Inform. Theory}, vol.~56, no.~5,
  May 2010.

\bibitem{Burnashev}
M.~V. Burnashev, ``Data transmission over a discrete channel with feedback.
  random transmission time,'' \emph{Probl. Inf. Transm.}, vol.~12, no.~4, pp.
  10--30, 1976.

\bibitem{Burnashev2}
------, ``Sequential discrimination of hypotheses with control of
  observations,'' \emph{Math. USSR, Izvestia}, vol.~15, no.~3, p. 419–440,
  1980.

\bibitem{Yamamoto}
H.~Yamamoto and K.~Itoh, ``Asymptotic performance of a modified
  {Schalkwijk-Barron} scheme for channels with noiseless feedback,''
  \emph{{IEEE} Trans. Inform. Theory}, vol.~25, no.~6, pp. 729--733, Nov. 1979.

\bibitem{Shulman}
N.~Shulman, ``Communication over an unknown channel via common broadcasting,''
  \emph{Ph.D. dissertation, Tel-Aviv Univ., Tel-Aviv}, vol.~25, no.~6, pp.
  729--733, Nov. 2003.

\bibitem{Draper}
S.~C. Draper, B.~J. Frey, and F.~R. Kschischang, ``Efficient variable length
  channel coding for unknown {DMC}s,'' in \emph{Proc. {IEEE} {ISIT}}, 2004.

\bibitem{Tchamkerten}
A.~Tchamkerten and E.~Telatar, ``A feedback strategy for binary symmetric
  channels,'' in \emph{Proc. {IEEE} {ISIT}}, Jun. 2002.

\bibitem{Tchamkerten2}
------, ``Optimal feedback schemes over unknown channels,'' in \emph{Proc.
  {IEEE} {ISIT}}, Jun. 2004.

\bibitem{Polyanskiy2}
Y.~Polyanskiy, H.~V. Poor, and S.~Verdu, ``Feedback in the non-asymptotic
  regime,'' \emph{{IEEE} Trans. Inform. Theory}, vol.~57, no.~8, pp.
  4903--4925, Aug. 2011.

\bibitem{Chen13}
T.~Chen, A.~R. Williamson, and R.~D. Wesel, ``Variable-length coding with
  feedback: Finite-length codewords and periodic decoding,'' in \emph{Proc.
  {IEEE} {ISIT}}, July 2013.

\bibitem{Shkim_GC}
S.~H. Kim, D.~K. Sung, and T.~Le-Ngoc, ``Performance analysis of incremental
  redundancy type hybrid {ARQ} for finite-length packets in {AWGN} channel,''
  in \emph{Proc. {IEEE} {GLOBECOM}}, Dec. 2013.

\end{thebibliography}
\newpage \thispagestyle{empty}
\end{document}